\documentclass[10pt]{article}
\usepackage{amsmath}
\usepackage{amssymb}
\usepackage[english]{babel}
\usepackage[latin1]{inputenc}
\usepackage{graphicx}
\usepackage{psfrag}
\usepackage{epsfig}
\newcommand{\nn}{\nonumber}
\numberwithin{equation}{section}
\title{\textbf{Covariant Impulse Approximation for the study of the internal structure of composite particles}}
\author{{\small Maurizio De Sanctis$^{1,2}$, Mario A. Acero$^2$, Diego A. Milanés$^2$, Carlos E. Sandoval$^2$}}
\date{\small{\textit{$^1$INFN, Sezione di Roma, $^2$ Universidad Nacional de Colombia}}}
\begin{document}
\maketitle%
\begin{abstract}
We present a brief review on the Impulse Approximation method to study processes of scattering off composite particles. We first construct the model in a non-relativistic fashion that enables us to extend the model to a covariant Impulse Approximation, which is needed for the study of high momentum transfer processes.
\end{abstract}
\section{Introduction}
The possibility of investigating the internal structure of composite
bound systems by means of the scattering processes is based on the
approximated validity of the following hypothesis, denoted as
Impulse Approximation (IA). The probing particle that makes a
scattering off the bound state, only interacts with a single
constituent at time. Moreover, the state of motion of the other
particles remains unaltered, that is their momenta do not change,
during the scattering process. For this reason, these particles
are called ``the spectators". This scheme can really work if the
interaction of the probing particle with the constituents of the
bound system is ``weak'', allowing for the perturbative treatment
of the scattering process, how it is assumed in the
previous description.
\\
In practice, the IA has been successfully applied to the study of
the electromagnetic and weak scattering processes. In particular,
very relevant physical results have been obtained studying the
elastic and inelastic form factors of the atomic nuclei, where a
non-relativistic treatment is sufficient to give a consistent
description of the scattering process.
\\
On the other hand, for the study of lepton scattering off
hadrons, where the non-relativistic model fails to reproduce the
experimental results, specially at high momentum transfer, a
generalization of the Impulse Approximation is required.
\\
Aim of the present work is to introduce, in a didactic way, a
model for the covariant generalization of the IA. More technical
details are discussed elsewhere \cite{DeSanc}. For simplicity and clarity we
shall explicitly treat two and three hadron composite systems, for
which accurate wave-functions can be constructed without great
difficulties. However, our model can be straightforwardly
generalized to the case of $A$ constituents.
%%%%%%%%%%%%%%%%%%%%%%%%%%%%%%%%%%%%%%%%%%%%%%%%%%%%%%%%%%%%%%%%%%%%%
\section{The non-relativistic model}
The general ideas of the previously introduced impulse
approximation have been used to define a non-relativistic
model, that we shall discuss in the present section with
the aim of fixing the starting point for its relativistic
generalization.
\\
Schematically, the Non Relativistic Impulse Aroximation NRIA, can be summarized by the following
equation:
\begin{equation}\label{eq:A}
\bar{A}_{FI} = \langle\Psi_F|\sum_{i=1}^A\widehat{O}_i
e^{i\vec{q}\vec{x}_i}|\Psi_I\rangle.
\end{equation}
The quantity $\bar{A}_{FI}$, that is the Transition Amplitude, or
better, the factor related to the bound state transition (being
the leptonic part a well known quantity), is defined as the
matrix-element, between the initial and final wave-functions of
the bound system $|\Psi_{F/I}\rangle$, of the transition operator
given by the sum of the transition operators of the single
particle (denoted by the letter $i$). The exponential operator
increases by the amount $\vec{q}$, the 3-momentum of the i-th
struck constituent, while the other constituents,
the spectators, do not change their momenta as required by the
hypothesis of the IA. This property takes the form
\begin{equation}\label{eq:mom}
e^{i\vec{q}\vec{x}_i}|\vec{p}_1,...,\vec{p}_i,...,\vec{p}_A\rangle
= |\vec{p}_1,...,\vec{p}_i+\vec{q},...,\vec{p}_A\rangle.
\end{equation}
The single particle ``current'' operator has
been synthetically denoted as $\widehat{O}_i$. Also, in equation
(\ref{eq:A}) we have neglected, for simplicity, all tensor
indices; for the same reason, isospin and/or charge operators have
not been explicitly indicated.
\\
In the non-relativistic quantum mechanics the wave function can be
easily separated into a first factor that describes the internal motion
and a plane-wave function of momentum $\vec{P}_{I/F}$ for the
motion of the composite particle as a whole. One has
\begin{equation}\label{eq:psi}
|\Psi_{F/I}\rangle = |\psi_{F/I}\rangle 
|\vec{P}_{F/I}\rangle.
\end{equation}
Furthermore, the single particle operators $\vec{x}_i$ can be
conveniently expressed in terms of the center of mass $\vec{R}$
and intrinsic coordinates. As anticipated, we shall treat
explicitly the two and three body cases for constituents of equal mass, 
but the procedure can be generalized to $A$ constituents without difficulties.
\\
The definition of the intrinsic coordinates is not unique. In the
following relations we report the most widely used choice, that
allows for highlighting the permutational symmetries of the wave
functions. For two particles we have:
\begin{eqnarray}\label{eq:coord2}
\vec r &=& \vec r_1 - \vec r_2,\nn \\
\vec R &=& \frac{1}{2}(\vec r_1 + \vec r_2),\nn \\
\vec p &=& \frac{1}{2}(\vec p_1 - \vec p_2), \\
\vec P &=& \vec p_1 + \vec p_2.\nn 
\end{eqnarray}
And for three particles:
\begin{eqnarray}\label{eq:coord3}
\vec{\rho} &=& \frac{1}{\sqrt{2}}(\vec r_1-\vec r_2),\nn \\
\vec{\lambda} &=& \frac{1}{\sqrt{6}}(\vec r_1+\vec r_2-2\vec r_3),\nn \\
\vec R &=& \frac{1}{3}(\vec r_1+\vec r_2+\vec r_3), \\ 
\vec p_{\rho} &=& \frac{1}{\sqrt{2}}(\vec p_1-\vec p_2),\nn \\
\vec p_{\lambda} &=& \frac{1}{\sqrt{6}}(\vec p_1+\vec p_2)-\sqrt{\frac{2}{3}}\vec p_3,\nn \\
\vec P &=& \vec p_1 + \vec p_2 + \vec p_3.\nn 
\end{eqnarray}
Furthermore, in the following we shall consider composite systems
of identical particles. For symmetry reasons the total
matrix-element $\bar{A}_{FI}$ may be written as $A$-times the
contribution of a single interacting constituent. For this we choose the
constituent number 2 for the two-body case ($A=2$) and the
constituent number 3 for the three-body case ($A=3$). Using 
the intrinsic operators of equations (\ref{eq:coord2}), (\ref{eq:coord3}), 
one obtain the conservation
of the total momentum as shown in the following equation. For
bound states of $k=2,3$ constituents we have
\begin{subequations}\label{eq:A2}
\begin{align}
\bar{A}_{FI} &= A_{FI}^{(k)}\delta^3(\vec{P}_F-\vec{P}_I-\vec{q})
\label{eqA2a}.
\end{align}
\end{subequations}
Also, the intrinsic matrix-element or vertex factor has the form
\begin{subequations}\label{eq:A2ab}
\begin{align}
A^{(2)}_{FI} &= 2\langle\psi_F^{(2)}|\widehat{O}_2
e^{-i\vec{q}\frac{\vec{r}}{2}}|\psi_I^{(2)}\rangle \label{eqA2b},\\
A^{(3)}_{FI} &= 3\langle\psi_F^{(3)}|\widehat{O}_3
e^{-i\vec{q}\lambda\sqrt{\frac{2}{3}}}|\psi_I^{(3)}\rangle \label{eqA2c},
\end{align}
\end{subequations}
for the two-and three-body cases, respectively.
\\
We explicitly consider the electric charge density transition
amplitudes that are obtained replacing in equations (\ref{eq:A2ab}) the generic operator $\widehat{O}_i$ with the
electric charge, that is
$$\widehat{O}_i = \widehat{e}_i,$$
with $i=2,3$ for equation (\ref{eqA2b}) and equation
(\ref{eqA2c}), respectively.
\\
For the intrinsic part of the electric current density transition amplitudes, one has
$$
\widehat{O}_2 e^{-i\vec{q}\frac{\vec{r}}{2}} \longrightarrow
-\frac{\hat{e}_2}{2m}\{\vec{p},e^{-i\vec{q}\frac{\vec{r}}{2}}\}
$$
for equation (\ref{eqA2b}), and
$$
\widehat{O}_3 e^{-i\vec{q}\vec{\lambda}\sqrt{\frac{2}{3}}} \longrightarrow
-\frac{\hat{e}_3}{2m}\sqrt{2}{3}\{\vec{p}_{\lambda},e^{-i\vec{q}\vec{\lambda}\sqrt{\frac{2}{3}}}\},
$$
for equation (\ref{eqA2c}). In the previous equations the
constituent particle mass $m$ has been introduced. A complete list of these electromagnetic current operators, including the relativistic corrections is given in ref. \cite{Foldy}.
\\
For the subsequent relativistic study, it is convenient to express
the typical vertex factors of equations (2.4 b,c) in terms of the
intrinsic momentum space wave-functions:
\begin{subequations}\label{eq:A-23}
\begin{align}
A_{FI}^{(2)}&=2\int d^3p \psi_F^{\dag}(\vec p -\frac{\vec q}{2})\hat O_2 \psi_I(\vec p),\\
A_{FI}^{(3)}&=3\int d^3p_{\rho} d^3p_{\lambda}\psi_F^{\dag}\left(\vec p_{\rho}, p_{\lambda}-\sqrt{2/3}\vec q\right)\hat O_3 \psi_I(p_{\rho}, p_{\lambda}).
\end{align}
\end{subequations}
Concluding this section, we point out that the model that we have
described is intrinsically non-relativistic. In fact, from a
formal point of view, the starting point of equation (\ref{eq:A})
cannot be put in a covariant form being absent in the exponential
operator any reference to the zero (time) components of the
position vectors $\vec x_i$ and of the momentum transfer $\vec
q$. Also, the separation of the intrinsic and CM variables of
equations (\ref{eq:coord2}), (\ref{eq:coord3}) is basically non-relativistic. 
For the electromagnetic interactions, it is possible to include in the
previous model all the relativistic corrections up to order
$c^{-2}$ (or $\frac{p^2}{m^2}$), by considering the Foldy
Wouthuysen expansion of the electromagnetic current matrix-elements
and adding the Lorentz boost corrections to the position and spin
operators. This procedure, that belongs to the theoretical
framework of the instant form relativistic quantum mechanics \cite{Dirac}, has
been applied to the study of the electromagnetic interactions of
the hadrons, obtaining numerically relevant results for some
low-momentum transfer observable, as the mean squared charge
radius of the proton in the constituent quark model \cite{DeSanc2}. However, it has not been found a viable method to generalize such procedure for
including the relativistic effects to all orders beyond the first
order corrections.
%%%%%%%%%%%%%%%%%%%%%%%%%%%%%%%%%%%%%%%%%%%%%%%%%%%%%%%%%%%%%%%%%%%%
%%%%%%%%%%%%%%%%%%%%%%%%%%%%%%%%%%%%%%%%%%%%%%%%%%%%%%%%%%%%%%%%%%%%%
\section{The relativistic model}
The NRIA described in the previous section, does not make use of
the ideas of relativity. In the following we show that it is
possible to define a relativistic three dimensional model, i.e.
without introducing the time components of the dynamical
quantities, conserving the main requirements of the Impulse
Approximation. For didactic reasons, we discuss the model in the
case of scalar constituents. However, the generalization to the
case of fermionic particles is straightforward, For the sake of
clarity we point out that we observe and describe the scattering
process in an ``observation reference frame'' -ORF-  that may be
for example the laboratory, the center of mass, the Breit, etc. On
the other hand we assume that the bound state can be represented
by a momentum space wave function, originally defined in its rest
frame, that will be denoted by an asterisk (*). Note that
whatever ORF we choose, the bound state will be in motion in the
initial and/or in the final state of the scattering process, so
that in the ORF, two different wave functions are observed, one
for the initial state and another for the final state. As in the
non-relativistic model, we make the hypothesis that only one
particle at time interacts with the external field.
\\ 
For identical particle systems, as before, we take the constituent 2
for the two-body case and the constituent 3 for the three body
case. The total result is then obtained multiplying by A=2 and
A=3, respectively. In the following we focus our attention on the
vertex factor $A_{FI}$. For the calculation of this quantity we
assume that the momenta of the spectators (i.e. the non
interacting constituents) remain unchanged during the scattering
process. As a specific assumption of the relativistic model, we
put the momenta of the spectators ``on the mass shell'', that is,
\begin{equation}\label{eq:mom1}
p_i=(E(\vec p_i),\vec p_i),\qquad \text{with}\qquad E(\vec p_i)=\sqrt{\vec p_i^2+m^2},
\end{equation}
or equivalently, in any reference frame
\begin{equation}\label{eq:mom2}
p_i^2=m^2,
\end{equation}
with $i=1$ and $i=1,2$ for the two-body and the three-body case
respectively. Note that $A-1$ momenta are required to describe the internal 
function of the bound system. Consequently, we use these momenta as the spatial
integration variables of the ORF for the calculation of the vertex
factor (i.e. matrix element) $A_{FI}$.We perform three
dimensional covariant integrations
\begin{equation}\label{eq:integ}
\int \frac{d^3 p_1}{E(\vec p_1)}, \qquad \int \frac{d^3
p_1}{E(\vec p_1)} \int \frac{d^3 p_2}{E(\vec p_2)},
\end{equation}
that ensure the covariance of the model. It means that, when the
ORF is changed, also the observable quantities $A_{FI}$ are
changed according to the Lorentz tensor properties of the
operators $\hat O_i$.
\\ 
Given that the wave functions are originally defined in the bound system rest reference frame we use
standard Lorentz transformations to determine the initial and
final reference frame momenta
\begin{equation}\label{eq:mom-IF}
p_{i I/F}^{*\alpha}=\Lambda_\mu^\alpha(\vec\beta_{I/F})p_i^\mu,
\end{equation}
where $\alpha=1,2,3$ denotes the spatial components of the rest
reference frame momenta and as before $i=1$ and $i=1,2$, for two- and three-body case. The Lorentz transformations depend on the velocity of the
initial/final state with respect to the ORF, that is
\begin{equation}\label{eq:beta}
\vec\beta_{I/F}=\frac{\vec P_{I/F}}{\sqrt{M_{I/F}^2+\vec P_{I/F}^2}}.
\end{equation}
Explicitly equation (\ref{eq:mom-IF}) takes the form
\begin{equation}\label{eq:Lorentz}
  \vec p_i^* = \vec p_i + \frac{\vec P}{M}\left[\frac{\vec P\cdot \vec p_i}{E+M} - E_i\right]
\end{equation}
Assuming that in the rest reference
frame the sum of the spatial momenta of the constituents is zero,
\begin{eqnarray}\label{eq:mom-zero}
\vec p_1^*+\vec p_2^*&=&0,\nonumber \\
\vec p_1^*+\vec p_2^*+\vec p_3^*&=&0,
\end{eqnarray}
we can express the Jacobi $A-1$ momenta $\vec p$ and $\vec p_{\rho}$, 
$\vec p_{\lambda}$ as linear combinations
of the $\vec p_{i I/F}^*$ determined by the Lorentz transformations
of equation (\ref{eq:Lorentz}), with, as usual, $i=1$ and $i=1,2$ for the two- and three-body case, respectively.
\\
We have almost all the ingredients to write down the vertex
factors of the covariant model. We shall discuss in the following
the
normalization factors that are introduced. We have
\\
\begin{equation}\label{eq:A2b}
A_{FI}^{(2)}=\frac{2}{J^{(2)}}\int\frac{d^3 p_1}{E(\vec p_1)}
\psi^\dag(\vec p^*_{1F})\sqrt{E(\vec p^*_{1F})}\hat
O_2\sqrt{E(\vec p^*_{1I})}\psi(\vec p^*_{1I}),
\end{equation}
\begin{equation}\label{eq:A3b}
A_{FI}^{(3)}=\frac{3}{J^{(3)}}\int\frac{d^3 p_1}{E(\vec p_1)}
\frac{d^3 p_2}{E(\vec p_2)}\psi^\dag(\vec p^*_{1F},\vec
p^*_{2F},)\sqrt{E(\vec p^*_{1F}E(\vec p^*_{2F})}\hat
O_3\sqrt{E(\vec p^*_{1I})E(\vec p^*_{2I})}\psi(\vec p^*_{1I},\vec
p^*_{2I}).
\end{equation}
The use of intrinsic variables $\vec p^*_{iI/F}$ instead of the Jacobi momenta,
makes necessary to introduce the normalization jacobians
\begin{subequations}\label{eq:J2}
\begin{align}
J^{(2)} &= 1 \\
J^{(3)} &= 3^{-3/2}
\end{align}
\end{subequations}
Furthermore, due to the use of the covariant integrations, it has been
necessary to introduce the invariant normalization factors
$$\sqrt{E(\vec p^*_{1I/F})},\qquad \sqrt{E(\vec p^*_{1I/F})E(\vec p^*_{2I/F})},$$
to ensure that in the static limit, \textit{i.e.} $\vec P_I=\vec P_F=0$, $M_F=M_I$,
the vertex factors reduce to the static mean value of the corresponding
operators $\hat O_2$ and $\hat O_3$.\\
This point may be examined in a slightly different way, shading some
light on the relativistic properties of the model. First, we observe
that the boost of the wave function is realized in the model by
expressing the intrinsic momenta $\vec p^*_i$ in terms of the ORF
momenta $\vec p_1$ and $\vec p_1$, $\vec p_2$ by means of the standard Lorentz transformation of equation (\ref{eq:Lorentz}).
\\
We can write
\begin{equation}\label{eq:psi}
\widetilde\psi(\vec p_i)=\psi(\vec p^*_i(\vec p_i)).
\end{equation}
However the boosted wave function $\widetilde\psi(\vec p_i)$ of the
previous equation is not normalized in a standard way, if one
integrates over the $\vec p_i$. Straightforward use of the
covariant integration rules show that
\begin{equation}\label{eq:jaco}
d^3p^*_i=d^3p_i\frac{E(\vec p^*_i)}{E(\vec p_i)}.
\end{equation}
In consequence, the correctly normalized boosted wave functions are
\begin{equation}\label{eq:psi-nor}
\psi(\vec p_1)=\left[\frac{E(\vec p^*_1)}{E(\vec
p_1)}\right]^{1/2}\psi(\vec p^*_1(\vec p_1)),
\end{equation}
\begin{equation}
\psi(\vec p_1,\vec p_2)=\left[\frac{E(\vec p^*_1)E(\vec
p^*_2)}{E(\vec p_1)E(\vec p_2)}\right]^{1/2}\psi(\vec p^*_1(\vec
p_1),\vec p^*_2(\vec p_2)),
\end{equation}\label{eq:psi12-nor}
to be integrated with respect to $d^3p_1$ and $d^3p_1d^3p_2$,
respectively. When a vertex factor is calculated, integrating with
respect to the same ORF momenta, the product of the denominators
of the previous equations give the factors $E(\vec p_1)$ and
$E(\vec p_1)E(\vec p_2)$, that are the denominators of the
covariant integration, while the numerators give the invariant
normalization factors thar have been previously introduced.
\\
This procedure of boosting the wave function, \textit{i.e.} by
transforming the on mass shell momenta, with the correct
normalization factors, represents a realization of the Point Form
Relativistic Quantum Mechanics \cite{DeSanc,Dirac}.
\\
In the model, the effects of the interaction that binds the
system, are only parametrically contained in the velocity
parameter $\vec\beta_ {I/F}$, while the momenta of the particles
are on mass shell.
\\
As a check of the consistency of the model we calculate the
non-relativistic limit of the vertex factor, showing that the same
result as the NRIA is obtained. First, we note that the factors
$$\frac{\sqrt{E(\vec p^*_{iF})E(\vec p^*_{iI})}}{E(\vec p_i)}$$
with $i=1$ and $i=1,2$, reduce to unity in the NR limit. The
vertex factors of equations (\ref{eq:A2b}) y (\ref{eq:A3b}) take
the form
\begin{equation}\label{eq:A2-ex}
A_{FI}^{(2)NR}=\frac{2}{J^{(2)}}\int d^3p_1\psi^\dag(\vec
p^{*NR}_{1F}) \hat O_2\psi(\vec p^{*NR}_{1I}),
\end{equation}
\begin{equation}\label{eq:A3-ex}
A_{FI}^{(3)NR}=\frac{3}{J^{(3)}}\int d^3p_1d^3p_2\psi^\dag(\vec
p^{*NR}_{1F},\vec p^{*NR}_{2F})\hat O_2\psi(\vec p^{*NR}_{1I},\vec
p^{*NR}_{2I}).
\end{equation}
Furthermore, the arguments of the wave functions that are $\vec
p_{iI/F}^{*NR}$, must be calculated by means of the NR limit of
the Lorentz transformations of equation (\ref{eq:Lorentz}), 
that are in other words the Galilean
transformations of the momenta, given in the following equation:
\begin{equation}\label{eq:galilean}
\vec{p}^*_{i(I/F)} = \vec{p}_i-\frac{\vec{P}_{I/F}}{A}.
\end{equation}
The reader should note that one could construct the NRIA model for
the vertex factors, by using the main hypotheses of the IA and the 
Galilean transformations of the momenta in equations (\ref{eq:A2-ex}) y
(\ref{eq:A3-ex}). Finally, standard change of variables according to equations (\ref{eq:coord2}) and (\ref{eq:coord3}) show that the same results as the originally defined NRIA are obtained, as given in equations (\ref{eq:A-23}).
%%%%%%%%%%%%%%%%%%%%%%%%%%%%%%%%%%%%%%%%%%%%%%%%%%%%%%%%%%%%%%%%%%%%%
%%%%%%%%%%%%%%%%%%%%%%%%%%%%%%%%%%%%%%%%%%%%%%%%%%%%%%%%%%%%%%%%%%%%%
\section{Conclusions}
In this paper, we have shown that it is possible to define a
relativistic generalization of the Impulse Approximation for the
scattering of leptonic particles on composite (hadronic) systems.
This model is deeply related to the Lorentz transformations
properties of the spatial variables and of the wave functions,
defining a specific form of Relativistic Quantum Mechanics (Point
Form). In a subsequent work we shall derive a conserved covariant
electromagnetic current, also considering the case of interacting
fermions.
%%%%%%%%%%%%%%%%%%%%%%%%%%%%%%%%%%%%%%%%%%%%%%%%%%%%%%%%%%%%%%%%%%%%%

\end{document}